%% file: main-no-formatting.tex
\documentclass[12pt]{article}

\usepackage{amsmath,amssymb,bbm}
\usepackage{graphicx,xcolor}
\usepackage{subfig}
\usepackage[ruled]{algorithm}
\usepackage{algpseudocode}
\usepackage{url}
\input{macros}

\usepackage{multirow}

\usepackage{natbib}
\usepackage[margin=1in]{geometry}

\newtheorem{assumption}{Assumption}

\renewcommand{\baselinestretch}{2}

\title{Semi-parametric Benchmark Dose Analysis with Monotone Additive Models}

\usepackage{authblk}
\author[1]{Alex Stringer}
\author[1]{Tugba Akkaya-Hocagil}
\author[1]{Richard Cook}
\author[2]{Louise Ryan}
\author[3]{Sandra W. Jacobson}
\author[3]{Joseph L. Jacobson}
\affil[1]{Department of Statistics and Actuarial Science, University of Waterloo, Waterloo, Canada}
\affil[2]{School of Mathematics and Statistics, University of Technology Sydney}
\affil[3]{Department of Psychiatry and Behavioural Neurosciences, Wayne State University School of Medicine}

\date{}

\counterwithout{algorithm}{subsection}

\begin{document}

\maketitle

\begin{abstract}
Benchmark dose analysis aims to estimate the level of exposure to a toxin that results in a clinically-significant adverse outcome
and quantifies uncertainty using the lower limit of a confidence interval for this level.
We develop a novel framework for benchmark dose analysis based on monotone additive dose-response models. 
We first introduce a flexible approach for fitting monotone additive models via penalized B-splines and Laplace-approximate marginal likelihood.
A reflective Newton method is then developed that employs de Boor's algorithm for computing splines and their derivatives for efficient estimation of the benchmark dose.
Finally, we develop and assess three approaches for calculating benchmark dose lower limits:
a naive one based on asymptotic normality of the estimator, one based on an approximate pivot,
and one using a Bayesian parametric bootstrap.
The latter approaches improve upon the naive method in terms of accuracy and are guaranteed to return a positive lower limit;
the approach based on an approximate pivot is typically an order of magnitude faster than the bootstrap, although they are both practically feasible to compute.
We apply the new methods to make inferences about the level of prenatal alcohol exposure associated with clinically significant
cognitive defects in children using data from an NIH–funded longitudinal study.
Software to reproduce the results in this paper is available at \url{https://github.com/awstringer1/bmd-paper-code}.
\end{abstract}

\input{01-introduction}

\input{02-benchmark-dosing}
\input{03-methods-dose}
\input{05-simulations.tex}

\input{06-dataanalysis.tex}

\input{07-discussion}

\bibliographystyle{apalike}
\bibliography{references.bib,Ref}

\appendix
\renewcommand{\subsection}{\Alph{section}}

\input{appendix-02-algorithms}

\end{document}

%% file: macros.tex
\newcommand{\response}{Y}
\newcommand{\cov}{x}
\newcommand{\covb}{\cov_{\texttt{b}}}

\newcommand{\intercept}{\alpha}

\newcommand{\covbest}{\widehat{\cov}_{\texttt{b}}}
\newcommand{\covlest}{\widehat{\cov}_{\texttt{l}}}

\newcommand{\covlestnormal}{\covlest^\Delta}
\newcommand{\covlestscore}{\covlest^\texttt{piv}}
\newcommand{\covlestboot}{\covlest^\texttt{boot}}
\newcommand{\covlestbayes}{\covlestboot}

\newcommand{\covmax}{\cov_{\texttt{max}}}
\newcommand{\covref}{\cov_{0}}
\newcommand{\refprob}{p_0}
\newcommand{\BMR}{p_{+}}
\newcommand{\threshold}{\tau_0}
\newcommand{\bmrconst}{c(\refprob,\BMR)}

\newcommand{\confint}{\widehat{\mathcal{C}}_n}

\newcommand{\conflevel}{\alpha}

\newcommand{\responsesd}{\sigma}
\newcommand{\responsesdest}{\widehat{\responsesd}}

\newcommand{\smoothfun}{f}
\newcommand{\smoothfunest}{\widehat{\smoothfun}}

\newcommand{\basisvec}{b}

\newcommand{\betavar}{\Sigma(\unconstrainedbasisweightvecmle)}
\newcommand{\betavaremp}{\widehat{\Sigma}(\unconstrainedbasisweightvecmle)}

\newcommand{\U}{U}
\newcommand{\Un}{\U_n}
\newcommand{\Unvar}{V_n}

\newcommand{\scorefunc}{\kappa_n}

\newcommand{\R}{\mathbb{R}}
\newcommand{\N}{\mathbb{N}}
\newcommand{\Normal}{\text{N}}
\newcommand{\iid}{\overset{\text{iid}}{\sim}}
\newcommand{\Tpose}{^{\textsf{T}}}
\newcommand{\inv}{^{-1}}

\newcommand{\PP}{\mathbb{P}}

\newcommand{\knot}{t}
\newcommand{\knots}{\boldsymbol{\knot}}

\newcommand{\responsedist}{\text{N}(0,1)}
\newcommand{\approxdist}{\overset{\cdot}{\sim}}

\newcommand{\confun}{g}
\newcommand{\confunest}{\widehat{\confun}}
\newcommand{\numconfuns}{m}
\newcommand{\concov}{z}

\newcommand{\numspline}{L}
\newcommand{\unconstrainedbasisweight}{\beta}
\newcommand{\unconstrainedbasisweightvec}{\boldsymbol{\unconstrainedbasisweight}}
\newcommand{\constrainedbasisweight}{\unconstrainedbasisweight^\texttt{c}}
\newcommand{\constrainedbasisweightvec}{\boldsymbol{\unconstrainedbasisweight}_\texttt{c}}
\newcommand{\unconstrainedbasisweightvecmle}{\widehat{\boldsymbol{\unconstrainedbasisweight}}}
\newcommand{\constrainedbasisweightvecmle}{\widehat{\boldsymbol{\unconstrainedbasisweight}}_\texttt{c}}

\newcommand{\constrainedbasisweightmle}{\widehat{\constrainedbasisweight}}

\newcommand{\constrainedbasisweightest}{\widehat{\constrainedbasisweight}}

\newcommand{\gambasisweight}{\gamma}
\newcommand{\gambasisweightvec}{\boldsymbol{\gambasisweight}}

\newcommand{\regparamvec}{\boldsymbol{\Psi}}
\newcommand{\paramdim}{D}
\newcommand{\varparamdim}{s}

\newcommand{\monodesignelem}{B}
\newcommand{\monodesignmat}{\boldsymbol{\monodesignelem}}
\newcommand{\gamdesignelem}{Z}
\newcommand{\gamdesignmat}{\boldsymbol{\gamdesignelem}}

\newcommand{\varparamvec}{\boldsymbol{\phi}}
\newcommand{\logprec}{\tau}
\newcommand{\logsmooth}{\lambda}
\newcommand{\logsmoothvec}{\boldsymbol{\logsmooth}}
\newcommand{\penaltymat}{\boldsymbol{S}}

\newcommand{\jnll}{\ell}
\newcommand{\mll}{\mathcal{L}}
\newcommand{\LAmll}{\mll_{\texttt{LA}}}

\newcommand{\norm}[1]{\lVert #1 \rVert}

\newcommand{\params}{\boldsymbol{\theta}}
\newcommand{\paramsmle}{\widehat{\params}}
\newcommand{\regparamvecmle}{\widehat{\regparamvec}}

\newcommand{\varparamvecmle}{\widehat{\varparamvec}}

\newcommand{\condhessian}{\boldsymbol{H}}

\newcommand{\order}{p}

%% file: 01-introduction.tex

\section{Introduction}

\subsection{Benchmark dose analysis}

Methodology for benchmark dose analysis is used by environmental toxicologists to quantify the level of exposure to a harmful substance associated with an adverse response. The US Environmental Protection Agency \citep{us_epa_benchmark_2012} and the European Food Safety Authority \citep{efsa_guidance_2022} use the lower limit of a $95\%$ confidence interval of the \emph{benchmark dose}---called the benchmark dose, lower (BMDL)---to set limits on acceptable levels of exposure to a wide variety of toxic substances.
In this paper we present a flexible and general computational and inferential framework for benchmark dose analysis and inference based on monotone additive models. We apply the new methodology to the problem of estimating levels of prenatal alcohol exposure that are associated with clinically significant cognitive defects in children.

\citet{crump_new_1984,crump_calculation_1995} introduced the concept of a benchmark dose (BMD) and recommended using the lower limit of a confidence interval for it (BMDL)
to define the acceptable exposure to a toxic substance.
Methods were developed first for parametric dose-response modeling of data obtained from designed experiments.
\citet{budtz-jorgensen_benchmark_2001} proposed a dose-response model based on a continuous exposure obtained from epidemiological (observational) studies that incorporates a propensity score-based adjustment for confounders. 
They derive an exact confidence interval for the BMD estimator based on a linear dose-response model with Gaussian errors. 
In recent work on parametric dose-response models, \citet{aerts_extended_2020} defined a family of parametric dose-response functions and recommended model-averaged point estimates based on a computationally intensive bootstrap procedure.

It has been argued that parametric dose-response modeling is not sufficiently flexible for the task of setting benchmark doses leading to the development of various semi- and non-parametric approaches.
\citet{wheeler_monotonic_2012} use Gibbs sampling to fit models based on monotone P-splines in a Bayesian framework that requires model-specific calculations, intensive computations, and manual convergence tuning and assessment. 
For binary outcomes, non-parametric methods were developed by \citet{piegorsch_nonparametric_2012,piegorsch_benchmark_2014} using isotonic regression for dose-response modelling, again relying on intensive bootstrap calculations for BMDL estimation; methods to deal with continuous responses were developed by \citet{lin_nonparametric_2015}. 
\citet{wheeler_quantile_2015} present a non-parametric quantile regression-based method.
These procedures are all characterized by their completely non-parametric dose-response models, and associated computationally-intensive inference procedures.

\subsection{Contributions}

We take a semi-parametric approach to benchmark dose analysis, combining the efficiency of the parameteric approaches with the flexibility of the non-parameteric methods.
Our approach uses monotone generalized additive models to estimate the benchmark dose and calculate the BMDL; the restriction to estimation of monotone dose-response
curves reflects the assumption that an increased exposure cannot yield a less adverse average response \citep{wheeler_monotonic_2012}.
We introduce a novel procedure for fitting monotone generalized additive models based on B-splines with coefficients parameterized to yield estimates of a monotone curve. 
Laplace-approximate marginal likelihood \citep{wood_fast_2011} is applied for smoothing parameter selection and uncertainty quantification \citep{wood_smoothing_2016}. 
We then develop a fast and stable method for solving the (random) non-linear equation defining the BMD estimate based on de Boor's algorithm for B-splines and their derivatives \citep{de_boor_practical_2001}, made feasible by deriving bounds on the estimated BMD and then applying a reflective Newton line search.
Finally, we introduce three methods for BMDL calculation: a) a ``Delta method'' lower limit based on asymptotic normality of the estimated BMD, b) inversion of a hypothesis test based on 
an approximate pivot obtained from
the estimating equation defining the BMD using another reflective Newton procedure, and c) a Bayesian parametric bootstrap based on asymptotic normality of the maximum likelihood estimator. 
The Delta method is fast but typically too conservative (coverage higher than nominal; see Table \ref{tab:simresultscovr} in Section \ref{sec:simulations}) 
and may give BMDLs that lie below zero exposure, providing no information about the BMD. 
The approximate pivot-based BMDL solves this problem, and is nearly as fast due to the development of a second de Boor/Newton algorithm. 
Further, the speed of the novel BMD estimation algorithm make the parametric bootstrap approach computationally feasible in typical applications.

\subsection{Motivating application}

Prenatal alcohol exposure (PAE) has been linked to a broad range of long-term cognitive and behavioral deficits \citep{jacobson_effects_2023}.  
However, there is relatively little information about the \textit{level} of PAE that is associated with clinically significant cognitive deficits.
Attempts to define what constitutes excessive drinking have been both qualitative \citep{Stratton1996, Hoyme2005, Chudley2005, Cook2016}
and quantitative \citep{Astley2000,Hoyme2016}, although \citet{Astley2000} emphasize that there is no  ``clear consensus on the amount of alcohol that can actually be toxic to the fetus".
Addressing this question has important clinical implications in terms of diagnosing children who have been adversely affected by pre-natal alcohol exposure.
We apply our method for semi-parameteric benchmark dose analysis to this problem using data from six United States National Institutes of Health-funded longitudinal
studies in which 
expectant women were interviewed regarding their drinking behaviour during pregnancy and their children followed through young adulthood
and assessed using a variety of cognitive tests.

%% file: 02-benchmark-dosing.tex

\section{Benchmark Dose Analysis and Monotone Splines}

\subsection{Benchmark dose analysis}\label{sec:bmdderivation}

Consider a response $\response_i\in\R$,
let $\cov_i\in\R$ represent the exposure for individual $i=1,\ldots,n$ in a sample of size $n$.
We consider the following dose-response model:
\begin{equation}\label{eqn:model}
    \response_i = \intercept +
    \smoothfun(\cov_i) + 
    \sum_{j=1}^{\numconfuns}\confun_j(\concov_{ij}) +
    \responsesd\epsilon_i, \
    \epsilon_i\iid\responsedist, \ 
    i=1,\ldots,n,
\end{equation}
where $\alpha$ is an intercept, $\smoothfun(\cov)$ is a strictly monotone decreasing
function of the exposure, and $\confun_j(\concov_j)$ are $\numconfuns$ functions of further covariates,
$\concov_j$, for $j=1,\ldots,\numconfuns$.
This specification assumes that an increase in $\cov$ reduces the mean response,
appropriate for settings where smaller values of the response are considered more adverse, as is the case with cognition scores.

We let $\covb$ represent the exposure level that yields a specified increase---called the benchmark response (BMR, denoted as $\BMR$)---in the probability of a response falling below a specified threshold; see \citet{crump_calculation_1995} and \citet{akkaya_hocagil_benchmark_2023} for detailed discussion of what follows. 
If $\refprob\in(0,1)$, let $\BMR\in(0,1-\refprob)$, and $\threshold(\covref,\concov)$ satisfy $\PP(\response < \threshold(\covref,\concov) ; \covref,\concov) = \refprob$ 
where $\PP(\cdot;\cov,\concov)$ is the probability distribution for $\response$ following the model (\ref{eqn:model}) with covariates $(\cov,\concov)$. 
Here $\covref\in\R$ is a baseline exposure and is almost always set at $\covref = 0$ in practice, representing an unexposed subject.
The benchmark dose (BMD, denoted as $\covb$) is defined as the value of $\cov$ satisfying
\begin{equation}\label{eqn:benchmarkdose-def}
    \PP(\response < \threshold(\covref,\concov) ; \covb,\concov) = \refprob+\BMR.
\end{equation}
Under the additive model (\ref{eqn:model}), $\covb$ is defined by the nonlinear equation $\U(\covb) = 0$,
where 
\begin{equation}\label{eqn:udef}
    \U(\cov) = \left(\smoothfun(\covref) - \smoothfun(\cov)\right)/\responsesd - \bmrconst,
\end{equation} 
and $\bmrconst = \Phi\inv(\refprob+\BMR) - \Phi\inv(\refprob)$ where $\Phi(\cdot)$ is the standard normal cumulative distribution function. 
It is clear from (\ref{eqn:udef}) that defined this way, $\covb$ does not depend on $\threshold(\covref,\concov)$ or $\concov$.
An estimate, $\covbest$, of $\covb$ solves the nonlinear equation $\Un(\covbest) = 0$, where
\begin{equation}\label{eqn:undef}
    \Un(\cov) = \left(\smoothfunest(\covref) - \smoothfunest(\cov)\right)/\responsesdest - \bmrconst,
\end{equation}
and $\smoothfunest(\cov),\responsesdest$ are estimates of $\smoothfun(\cov)$ and $\responsesd$.

\subsection{Quantifying uncertainty in the benchmark dose}\label{sec:inferencebenchmarkdose}

To address the uncertainty in estimation of the benchmark dose it is common to base guidelines on a lower limit of 
the benchmark dose (or benchmark dose-lower; BMDL),
defined as the lower endpoint of an approximate $95\%$ confidence interval for $\covb$. 
\citet{budtz-jorgensen_benchmark_2001} derive the exact sampling distribution of $\covbest$ in linear dose-response models and use it to give a formula for a BMDL.
The explicit sampling distribution of $\covbest$ is not available in more flexible dose-response models, and a common approach is to make inferences based on the Delta method,
relying on asymptotic normality of $\covbest$.
Computationally-intensive bootstrap-based BMDL estimation is also possible; see \citet{moerbeek_comparison_2004} for a review of existing methods.
We derive these methods, as well as one additional method for uncertainty quantification of the BMD based on semi-parametric dose-response models in Section \ref{sec:computationbmdl}.

\subsection{Monotone B-splines}\label{sec:monotonesplines}

The dose-response function is represented as a monotone spline function, which proves useful for both fitting the dose-response model and rapid computation of the BMD and BMDL,
and also ensures that a BMD exists for some choice of the BMR; see Section \ref{sec:inference}. 
The function $\smoothfun(\cov)$ can be written as 
\begin{equation}\label{eqn:bsplinefunc}
\smoothfun(\cov) = \sum_{l=1}^{\numspline}\basisvec_{l,\order}(\cov)\unconstrainedbasisweight_l,
\end{equation}
where $\basisvec_{l,\order}(\cov)$ is the $l^{th}$ B-spline function of order $\order$ on the interval $[a,b]\subset\R$ with knots $\knots = (\knot_1,\ldots,\knot_{\numspline+\order})$ such that $a \leq \knot_1\leq\cdots\leq\knot_{\numspline+\order}\leq b$. 
Using cubic B-splines having $\order=4$, we follow the standard practice of choosing a large value of $\numspline$
and controlling over-fitting through penalized estimation; see \citet{wood_smoothing_2016,wood_p-splines_2017} for details.

Note that B-splines are defined recursively with $\basisvec_{l,1}(x) = I(\left[\knot_l,\knot_{l+1}\right))$ and $\basisvec_{l,p}(x) = \omega_{l,p-1}\basisvec_{l,p-1}(x) + (1-\omega_{l+1,p-1})\basisvec_{l+1,p-1}(x)$ where $\omega_{l,p}(x) = (x-\knot_l)/(\knot_{l+p}-\knot_l) I(\knot_{l+p}\neq\knot_l)$. These definitions are used during model fitting (Section \ref{sec:modelfitting}) to compute the design matrix where the $\cov$ at which the splines are to be evaluated are specified in advance.

The numerical methods of Section \ref{sec:inference} (see also Algorithms \ref{alg:newtonreflect} and \ref{alg:newtonscore} in Appendix \ref{app:algorithms}) require computation of the entire spline function (\ref{eqn:bsplinefunc}) within an iterative root-finding procedure at many values of $\cov$ that cannot be known in advance. 
In this setting, computing $\basisvec_{1,p}(x),\ldots,\basisvec_{\numspline,p}(x)$ naively and computing $\smoothfun(\cov)$ using (\ref{eqn:bsplinefunc}) is too computationally intensive, since this wastefully ignores the fact that only $\order+1\ll\numspline$ of the basis functions are non-zero for each $x$. 
de Boor's algorithm \citep{de_boor_practical_2001} computes the entire spline function $\smoothfun(\cov)$ efficiently by avoiding computation of the individual basis functions $\basisvec_{l,p}(\cov)$, and ignoring computations known to add zero to the overall sum; see Algorithm \ref{alg:deboor} in Appendix \ref{app:algorithms}. 
This greatly increases the speed of the computations involved in estimating the BMD and hence makes the parametric bootstrap of Section \ref{sec:computationbmdl} practiclly feasible.

The derivative of a B-spline basis function is given by $\basisvec^\prime_{l,\order}(\cov) = (\order-1)(\basisvec^\prime_{l,\order-1}(\cov)/(\knot_{l+\order-1}-\knot_l) - \basisvec^\prime_{l+1,\order-1}(\cov)/(\knot_{l+\order}-\knot_{l+1}))$, and the derivative of the entire spline function
(\ref{eqn:bsplinefunc}) is:
\begin{equation}\label{eqn:bsplinederiv}
\smoothfun^\prime(\cov) = (\order-1)\sum_{l=2}^{\numspline}\basisvec_{l,\order-1}(\cov)\frac{\unconstrainedbasisweight_{l} - \unconstrainedbasisweight_{l-1}}{\knot_{l+\order-1}-\knot_l}.
\end{equation}
From this it can be seen that $\unconstrainedbasisweight_{1} < \cdots < \unconstrainedbasisweight_{\numspline}$ is a sufficient condition for $\smoothfun(\cov)$ to be strictly monotone decreasing, and this is straightforward to ensure during model fitting via re-parameterization as introduced by \citet{pya_shape_2015}; see Section \ref{sec:modelfitting}. Further, because the derivatives of a spline function are also spline functions (linear combinations of spline basis functions), de Boor's algorithm is also applied to obtain the derivatives required to implement Newton's method, leading to further computational benefits.

%% file: 03-methods-dose.tex

\section{Inferences about the Benchmark Dose}\label{sec:inference}

\subsection{Existence of the benchmark dose}\label{sec:statisticalprobpertiesdose}

The benchmark dose $\covb$, defined as the solution to $\U(\cov)=0$ where $\U(\cov)$ is given by (\ref{eqn:benchmarkdose-def}), is not guaranteed to exist for every choice of $\covref,\refprob,\BMR$. 
The specification of $\covref,\refprob,\BMR$ should be context-dependent; see \citet{haber_benchmark_2018}. 
We repeat that $\covref=0$ is almost always specified, representing an unexposed subject.


Assumption \ref{assn:monotone} is sufficient to ensure that for some $p_0,\BMR$, there exists a value of $\covb$ satisfying $\U(\covb)=0$:

\begin{assumption}\label{assn:monotone}
\emph{(a)} The unknown function $\smoothfun(\cov)$ is continuously differentiable with \\ $\sup_{\cov\in\R}|\smoothfun^\prime(\cov)|<\infty$ and $\smoothfun^\prime(\cov)<0$ for every $\cov\in\R$; 
and \emph{(b)} with probability 1 the estimated function $\smoothfunest(\cov)$ is continuously differentiable with $\sup_{\cov\in\R}|\smoothfunest^\prime(\cov)|<\infty$ and $\smoothfunest^\prime(\cov)<0$ for every $\cov\in\R$.
\end{assumption}
Assumption \ref{assn:monotone} (a) states that higher exposure cannot be associated with an equal or better expected response. 
Assumption \ref{assn:monotone} (b) can be satisfied by applying a monontonic smoother, as we do in Sections \ref{sec:monotonesplines} and \ref{sec:modelfitting}.
Since $\U(\covref) < 0$ from Assumption \ref{assn:monotone}, by the intermediate value theorem $\U(\covmax)>0$ is a sufficient condition for the existence of $\covb\in(\covref,\covmax)$ satisfying $\U(\covb)=0$. 
Further, a sufficient condition for the estimate $\covbest$ to exist is $\Un(\covmax)>0$. In practice, we recommend to check $\Un(\covmax)>0$, and decrease $\BMR$ if this condition is not satisfied. In cases where the underlying dose-response curve is not monotonic, a benchmark dose may not exist, and this is important to consider in applications.


\subsection{Fitting the dose-response model}\label{sec:modelfitting}

Under the B-spline representation (\ref{eqn:bsplinefunc}), the dose-response model (\ref{eqn:model}) is:
\begin{equation}\label{eqn:modelbspline}
    \response_i = \intercept +
    \sum_{l=1}^{\numspline}\basisvec_{l,4}(\cov_i)\constrainedbasisweight_l + 
    \sum_{j=1}^{\numconfuns}\sum_{l=1}^{\numspline}\basisvec_{l,4}(\concov_{ij})\gambasisweight_{jl} +
    \responsesd\epsilon_i, \
    \epsilon_i\iid\responsedist, \ 
    i=1,\ldots,n.
\end{equation}
Monotonicity can be enforced by reparameterization \citet{pya_shape_2015}.
We set $\constrainedbasisweight_1 = \unconstrainedbasisweight_1$ and $\constrainedbasisweight_l = \constrainedbasisweight_{l-1} - \exp\left(\unconstrainedbasisweight_l\right),l=2,\ldots,\numspline$, which guarantees that $\constrainedbasisweight_{1} < \cdots < \constrainedbasisweight_{\numspline}$ and hence that $\smoothfun(\cov)$ and its estimate, $\smoothfunest(\cov) = \basisvec_{1,4}(\cov)\constrainedbasisweightest_1 + \cdots + \basisvec_{\numspline,4}(\cov)\constrainedbasisweightest_\numspline$, are monotone decreasing.
The unknown parameters to be estimated are $\regparamvec = (\alpha,\unconstrainedbasisweightvec,\gambasisweightvec)\in\R^\paramdim$ where $\alpha\in\R,\unconstrainedbasisweightvec = (\unconstrainedbasisweight_1,\ldots,\unconstrainedbasisweight_\numspline)\in\R^\numspline,$ and $\gambasisweightvec = (\gambasisweight_{11},\ldots,\gambasisweight_{\numconfuns\numspline})\in\R^{\numconfuns\numspline}$, so that $\paramdim = 1 + \numspline(1 + \numconfuns)$.

We define the design matrices $\monodesignmat = \left(\monodesignelem_{il}\right)$ with $\monodesignelem_{il} = \basisvec_{l,4}(\cov_i)$ and $\gamdesignmat = \left[\gamdesignmat_1 : \cdots : \gamdesignmat_\numconfuns\right]$ with $\gamdesignmat_{j} =  \left(\gamdesignelem_{jil}\right)$ and $\gamdesignelem_{jil} = \basisvec_{l,4}(\concov_{ij})$, and define $\varparamvec = (\logprec,\logsmoothvec) \in\R^{\varparamdim}$ where $\varparamdim = \numconfuns+2,\logprec = -2\log\sigma$, and $\logsmoothvec = (\logsmooth_1,\ldots,\logsmooth_{\numconfuns+1})$ where $\logsmooth_j\in\R$ are (log) smoothing penalty parameters to be estimated. Finally, let $\penaltymat_1,\penaltymat_2,\ldots,\penaltymat_{\numconfuns+1}$ be matrices of integrated squared second B-spline derivatives, computed according to the algorithm of \citet{wood_p-splines_2017}, and let $\penaltymat_{\logsmoothvec} = \text{diag}\left(e^{\logsmooth_2}\penaltymat_2,\ldots,e^{\logsmooth_{\numconfuns+1}}\penaltymat_{\numconfuns+1}\right)$. 
A penalized joint (negative) log-likelihood for the unknown parameters, $(\regparamvec,\varparamvec)$, is
\begin{equation}
    \jnll(\regparamvec,\varparamvec) = \frac{1}{2}e^\logprec\norm{\boldsymbol{y} - \alpha\boldsymbol{1}_n - \monodesignmat\constrainedbasisweightvec - \gamdesignmat\gambasisweightvec}^2_2 + e^{\logsmooth_1}\unconstrainedbasisweightvec\Tpose\penaltymat_1\unconstrainedbasisweightvec + \gambasisweightvec\Tpose\penaltymat_{\logsmoothvec}\gambasisweightvec,
\end{equation}
and a marginal likelihood for $\varparamvec$ is
\begin{equation}
\mll(\varparamvec) = \int\exp\left\{ -\jnll(\regparamvec,\varparamvec)\right\}d\regparamvec.
\end{equation}
We define the profile maximum likelihood estimator as $\regparamvecmle(\varparamvec) = \text{argmin}_{\regparamvec}\jnll(\regparamvec,\varparamvec)$, and the Hessian matrix $\condhessian(\varparamvec) = -\partial^2_{\regparamvec}\jnll(\regparamvecmle(\varparamvec),\varparamvec)$. 
Inferences about $\varparamvec$ are to be based on the Laplace-approximate maximum marginal likelihood estimator, $\varparamvecmle = \text{argmax}_{\varparamvec}\LAmll(\varparamvec)$, where
\begin{equation}
\LAmll(\varparamvec) = \left(2\pi\right)^{\paramdim/2}\lvert\condhessian(\varparamvec)\rvert^{-1/2}\exp\left\{ -\jnll(\regparamvecmle(\varparamvec),\varparamvec)\right\}
\end{equation}
is a Laplace approximation to $\mll(\varparamvec)$. 
Point estimates of $\regparamvec$ are then given as $\regparamvecmle(\varparamvecmle)$, and a point estimate of the unknown dose-response function at any $\cov\in\R$ is $\smoothfunest(\cov) = \basisvec(\cov)\Tpose\constrainedbasisweightvecmle$ with $\basisvec(\cov) = (\basisvec_{1,4}(\cov),\ldots,\basisvec_{\numspline,4}(\cov))\Tpose$. 
Estimates $\confunest_j$ for each $\confun_j,j=1,\ldots,\numconfuns$ are analogously obtained.
Uncertainty quantification is discussed in Section \ref{sec:computationbmdl}.

Spline calculations are carried out using the \texttt{Rcpp} framework \citep{eddelbuettel_rcpp_2011}.
The Template Model Builder (TMB) framework \citep{kristensen_tmb_2016} is used for automatic differentiation and Laplace approximation, yielding efficient computation of $\log\LAmll(\varparamvec)$ and $\partial_{\varparamvec}\log\LAmll(\varparamvec)$.
We compute $\varparamvecmle$ via quasi-Newton optimization based on these quantities. 
The linear constraints $\sum_{i=1}^n\smoothfunest(\cov_i) = 0$ and $\sum_{i=1}^n\confunest(\concov_{ij}) = 0,j=1,\ldots,\numconfuns$ are imposed directly, to avoid ``constraint-absorbing'' reparameterizations \citep{stringer_identifiability_2023}. 
Note that because $\Un(\cdot)$ is invariant to the addition of constant terms to $\smoothfunest$, estimation of $\covb$ is invariant to the choice of linear constraints. However, such constraints are required to ensure the identifiability of $\alpha$ as well as more than one $\smoothfun$ and $\confun$ together, so they still must be used.

\subsection{Computational method for benchmark dose estimation}\label{sec:numericalmethods}

Given a fitted dose-response model, we obtain the estimate, $\covbest$, defined as the solution to a non-linear equation, $\Un(\covbest)=0$, via a reflective Newton line search. The efficiency and stability of the method is critically important in the parametric bootstrap of Section \ref{sec:computationbmdl} where we repeat the procedure thousands of times to calculate a benchmark dose lower limit. 
Efficient computation of the required B-spline functions and derivatives to implement the Newton iteration is facilitated by de Boor's algorithm,
and a stable iteration is achieved by bounding the solution to a known interval.

Bounds on $\covbest$ are obtained by noting that Assumption \ref{assn:monotone} (b) guarantees that $\Un(\cdot)$ is continuous and strictly monotonic, and further that $\Un(\covref) = -\bmrconst < 0$. 
We therefore check the condition that $\Un(\covmax) > 0$ in applications.
If this constraint is satisfied, then by definition of $\covbest$ and the intermediate value theorem, $\covref<\covbest<\covmax$.
We then augment the typical Newton iteration with the reflective transformation given by \citet{coleman_convergence_1994}, using $(\covref,\covmax)$ as the required interval within which the solution is known to lie.
The full algorithm is given in Algorithm \ref{alg:newtonreflect} in the Appendix, which depends further on de Boor's algorithm (Algorithm \ref{alg:deboor}). 



\subsection{Computation of benchmark dose lower limits}\label{sec:computationbmdl}

Inferences about $\covb$ are based on a benchmark dose lower limit, $\covlest$, defined as the lower endpoint of a $95\%$ confidence interval for $\covb$; see Section \ref{sec:inferencebenchmarkdose}. 
Here we introduce three candidate lower limits: one based on a Bayesian parametric bootstrap, one based on the Delta method for $\covbest$, and one based on an approximate pivot obtained from $\Un$.

\subsubsection{Bayesian parametric bootstrap}\label{subsubsec:bootstrapbmdl}

To quantify uncertainty in the estimated regression coefficients from the dose-response model, $\regparamvecmle$, and functions of them including the estimated BMD, $\covbest$, we consider inference based on (approximate) posterior samples in a Bayesian framework; see \citet{wood_smoothing_2016} for a detailed exposition. For notational clarity, denote the entire vector of spline weights and variance/smoothing parameters by $\params = (\regparamvec,\varparamvec)$, with estimate $\paramsmle = (\regparamvecmle,\varparamvecmle)$. 
Given $\paramsmle$, we employ the approximation $\params | \boldsymbol{\response} \sim \text{N}(\paramsmle,\condhessian\inv(\varparamvecmle))$. We draw samples, $\{\params_j\}_{j=1}^{M}$ where $M\in\N$, from $\params | \boldsymbol{\response}$ using the method of \citet{rue_fast_2001}, based on the Cholesky decomposition of $\condhessian(\varparamvecmle)$.

Given $\params_j$, a sample, $\regparamvec_j$, from $\regparamvec|\boldsymbol{\response}$ is obtained by indexing the first $\text{dim}(\regparamvec)$ components of $\params_j$; a sample, $\varparamvec_j$, from $\varparamvec|\boldsymbol{\response}$ is obtained as the remaining $\text{dim}(\varparamvec)$ components. 
Further, a sample, $\covb^j$, from $\covb|\boldsymbol{\response}$ is obtained by running Algorithm \ref{alg:newtonreflect} with $\params_j$ as the input.
The bootstrap BMDL, $\covlestbayes$, is the $2.5^{th}$ percentile of the $M$ BMD samples obtained in this manner. 
This procedure involves running Algorithm \ref{alg:newtonreflect} $M$ times, and is made computationally feasible by the use of de Boor's algorithm for both $\smoothfunest$ and $\smoothfunest^\prime$; see Algorithm \ref{alg:newtonreflect}. 
A more naive parametric bootstrap would fit the entire dose-response model (\ref{eqn:modelbspline}) $M$ times, 
but the use of (approximate) posterior samples requires only a single model fit, and hence is computationally efficient.

\subsubsection{Delta method}\label{subsubsec:deltabmdl}

Although the bootstrap procedure of Section \ref{subsubsec:bootstrapbmdl} is fast, a faster Delta-method lower bound is obtained from the frequentist asymptotic approximation:
\begin{equation}
    \regparamvecmle \overset{\cdot}{\sim} \text{N}\left(\regparamvec,\condhessian\inv(\varparamvecmle)\right).
\end{equation}
From this, we obtain $\Un(\covb)/\Unvar(\covb)^{1/2}\approxdist\text{N}(0,1)$, where
$$
\Unvar(\covb) = \text{Var}\{\Un(\covb)\} = \frac{1}{\responsesdest^2}\left(\basisvec(\covref)-\basisvec(\covb)\right)\Tpose\betavar\left(\basisvec(\covref)-\basisvec(\covb)\right),
$$
and $\betavar = \text{Cov}(\unconstrainedbasisweightvecmle)$.
A linear Taylor series approximation, $\Un(\covb)\approx\Un^{\prime}(\covbest)(\covb-\covbest)$, gives:
\begin{equation}\label{eqn:approxxb}
\frac{\Un^{\prime}(\covbest)}{\Unvar(\covbest)^{1/2}}\left(\covbest - \covb\right)\approxdist \Normal\left(0,1\right).
\end{equation}
We obtain samples, $\constrainedbasisweightvec^j$, from $\params_j$ by indexing out $\unconstrainedbasisweightvec^j$ from each and then applying
the transformation given in Section \ref{sec:modelfitting}.
We use the sample covariance matrix of $\constrainedbasisweightvec^1,\ldots,\constrainedbasisweightvec^M$, $\betavaremp = \text{Cov}(\constrainedbasisweightvec^1,\ldots,\constrainedbasisweightvec^M)$, to estimate $\betavar$.
We then compute $\Unvar(\covbest)$ directly, using the recursions for B-splines.
A Delta method-based BMDL is then:
\begin{equation}
\covlestnormal = \covbest - 2\frac{\Unvar(\covbest)^{1/2}}{|\Un^{\prime}(\covbest)|}.
\end{equation}
Computation of $\Un^{\prime}(\covbest)$ via de Boor's algorithm is described within Algorithm \ref{alg:newtonreflect}.


\subsubsection{Approximate pivot}\label{subsubsec:pivotbmdl}

The Delta method lower bound of Section \ref{subsubsec:deltabmdl} is fast to compute, but may be inaccurate and return a lower bound, $\covlestnormal<\covref$, in practice. Such a lower bound provides no information about $\covb$.
We observe such an unusable estimate in a significant proportion of simulations (Table \ref{tab:simresultscomp} in Section \ref{sec:simulations}); in the application to the PAE study data in Section \ref{sec:paeanalysis},
we find that $\covlestnormal$ is very close to zero and substantially smaller than the other two BMDLs.
We address this problem by introducing the approximate pivot-based BMDL, $\covlestscore \in \inf\confint(\conflevel)$, where
$$
\confint(\conflevel) = \left\{\cov\in\R: \Un(\cov)^2  < \Unvar(\cov)\chi^{2}_{1,\conflevel} \right\}.
$$
This is motivated by the approximate pivot $\Un(\covb)^2/\Unvar(\covbest)\approxdist\chi^2_1$ which follows immediately from $\Un(\covb)/\Unvar(\covb)^{1/2}\approxdist\text{N}(0,1)$.
However, we do not employ a Delta method, instead relying directly on the approximate distribution of the pivot.
This yields accurate intervals that cannot cross zero; see the simulations in Section \ref{sec:simulations}.

Computation of $\covlestscore$ is more involved than that of $\covlestnormal$. We utilize another application of the reflective Newton line search coupled with de Boor's algorithm, yielding a procedure nearly as fast as the Delta method and much faster than the bootstrap, as follows.
Define the function $\scorefunc(\cov) = \Un(\cov)^2  - \Unvar(\cov)\chi^{2}_{1,\conflevel}$ and note that $\covlestscore\in\inf\left\{\cov:\scorefunc(\cov)=0\right\}$. Accordingly, we use an appropriate modification of Algorithm \ref{alg:newtonreflect} to find an appropriate zero of $\scorefunc$. The required derivative is $\scorefunc^\prime(\cov) = 2\Un(\cov)\Un^\prime(\cov) - \Unvar^\prime(\cov)\chi^{2}_{1,\conflevel}$ where $\Unvar^\prime(\cov) = -2\basisvec^\prime(\cov)\Tpose\betavar\left(\basisvec(\covref)-\basisvec(\cov)\right)$. Both the B-spline basis function vectors, $\basisvec(\cov)$, and their vectors of derivatives, $\basisvec^\prime(\cov)$, are computed directly using the recursions described in Section \ref{sec:monotonesplines}. Bounds are obtained by observing that $\scorefunc(\covref) = \bmrconst^2>0$ and $\scorefunc(\covbest) = - \Unvar(\cov)\chi^{2}_{1,\conflevel}<0$, and applying the intermediate value theorem to conclude that there exists $\covlestscore\in(\covref,\covbest)$. 
Although such a root may or may not not be unique, we have not observed problems with convergence or stability in our experiments (Section \ref{sec:simulations}) or data analysis (Section \ref{sec:paeanalysis}). The full algorithm is given in Algorithm \ref{alg:newtonscore} in Appendix \ref{app:algorithms}.

%% file: 05-simulations.tex

\section{Empirical Performance and Computational Considerations}\label{sec:simulations}

We compare via simulations the empirical performance of the three candidate BMDL's: $\covlestbayes$ (Section \ref{subsubsec:bootstrapbmdl}), $\covlestnormal$ (Section \ref{subsubsec:deltabmdl}), and $\covlestscore$ (Section \ref{subsubsec:pivotbmdl}).
All computations are performed using the \texttt{semibmd} \texttt{R} package which implements the methods described in Section \ref{sec:inference}.
We report empirical coverage probabilities and relative computation times for $100,000$ replicates simulated under the true dose-response function $\smoothfun(\cov) = \exp(-s\cov)$ for varying $s>0$. This gives a monotone curve of varying steepness, where smaller $s$ yields a flatter curve and hence a more difficult estimation problem.
Table \ref{tab:simresultscovr} shows the bias and empirical coverage rates of $\covlestnormal,\covlestscore$, and $\covlestbayes$, the latter based on $1000$ parametric bootstrap replications for each of the $100,000$ simulated replicates; 
since these are lower limits of two-sided $95\%$ confidence intervals, they are expected to yield coverage of approximately $97.5\%$. 
The performance of $\covlestnormal$ is broadly unsatisfactory: the average coverage is too high in all cases, which includes a substantial proportion of computed lower limits having $\covlestnormal<\covref$; see Table \ref{tab:simresultscomp}. 
Note that $\covlestscore$ and $\covlestbayes$ are within 2 standard errors of the nominal coverage in all cases, with the former occasionally slightly lower and the latter occasionally slightly higher.
{\renewcommand{\baselinestretch}{1}
\begin{table}[p]
\centering
\begin{tabular}{ccc|rrrr}
$n$ & $s$ & $\sigma$ & EBias & $\%$ECP($\covlestnormal$) & $\%$ECP($\covlestscore$) & $\%$ECP($\covlestbayes$) \\
\hline
\multirow{12}{*}{$200$}  & \multirow{3}{*}{$.1$} & $.1$ &   5.7(0.06) & 100.0(0.00) &  94.6(0.07) & 100.0(0.00) \\
                         &                       & $.2$ &  -7.9(0.11) & 100.0(0.00) &  94.7(0.08) & 100.0(0.00) \\
                         &                       & $.5$ & -91.8(0.14) & 100.0(0.00) & 100.0(0.00) & 100.0(0.00) \\
                         & \multirow{3}{*}{$.5$} & $.1$ &   0.7(0.01) &  99.8(0.01) &  97.3(0.05) &  99.1(0.03) \\
                         &                       & $.2$ &   3.0(0.03) &  99.9(0.01) &  97.0(0.05) &  99.6(0.02) \\
                         &                       & $.5$ &   5.7(0.07) & 100.0(0.00) &  93.4(0.08) & 100.0(0.00) \\
                         & \multirow{3}{*}{$1$}  & $.1$ &   0.2(0.00) &  99.5(0.02) &  97.0(0.05) &  98.5(0.04) \\
                         &                       & $.2$ &   1.2(0.02) &  99.5(0.02) &  96.6(0.06) &  99.1(0.03) \\
                         &                       & $.5$ &   5.4(0.04) & 100.0(0.01) &  96.1(0.06) &  99.9(0.01) \\
                         & \multirow{3}{*}{$2$}  & $.1$ &   0.7(0.02) &  98.0(0.04) &  95.6(0.06) &  97.9(0.05) \\
                         &                       & $.2$ &   1.5(0.03) &  98.1(0.04) &  95.8(0.06) &  98.8(0.03) \\
                         &                       & $.5$ &   2.8(0.02) & 100.0(0.01) &  96.5(0.06) &  99.6(0.02) \\
                         & \multirow{3}{*}{$5$}  & $.1$ &   0.1(0.01) &  98.3(0.04) &  95.9(0.06) &  97.4(0.05) \\
                         &                       & $.2$ &   0.2(0.01) &  99.5(0.02) &  96.7(0.06) &  98.5(0.04) \\
                         &                       & $.5$ &   1.2(0.01) & 100.0(0.01) &  95.9(0.06) &  99.1(0.03) \\
\hline
\multirow{12}{*}{$500$}  & \multirow{3}{*}{$.1$} & $.1$  &   3.8(0.04) &  99.8(0.02) &  97.4(0.05) & 100.0(0.01) \\
                         &                       & $.2$  &  -4.1(0.08) & 100.0(0.00) &  96.0(0.07) & 100.0(0.00) \\
                         &                       & $.5$  & -87.5(0.12) & 100.0(0.00) & 100.0(0.00) & 100.0(0.00) \\
                         & \multirow{3}{*}{$.5$} & $.1$  &   0.2(0.00) &  99.6(0.02) &  97.2(0.05) &  98.6(0.04) \\
                         &                       & $.2$  &   1.4(0.02) &  99.9(0.01) &  97.3(0.05) &  99.2(0.03) \\
                         &                       & $.5$  &   4.4(0.05) &  99.8(0.01) &  97.0(0.06) & 100.0(0.00) \\
                         & \multirow{3}{*}{$1$}  & $.1$  &   0.1(0.00) &  99.0(0.03) &  96.9(0.05) &  98.2(0.04) \\
                         &                       & $.2$  &   0.4(0.01) &  99.6(0.02) &  97.1(0.05) &  98.7(0.04) \\
                         &                       & $.5$  &   3.0(0.03) &  99.9(0.01) &  97.0(0.05) &  99.6(0.02) \\
                         & \multirow{3}{*}{$2$}  & $.1$  &   0.1(0.01) &  98.4(0.04) &  96.5(0.06) &  97.8(0.05) \\
                         &                       & $.2$  &   0.8(0.02) &  98.4(0.04) &  96.3(0.06) &  98.5(0.04) \\
                         &                       & $.5$  &   1.8(0.02) &  99.5(0.02) &  96.5(0.06) &  99.2(0.03) \\
                         & \multirow{3}{*}{$5$}  & $.1$  &   0.0(0.00) &  98.1(0.04) &  96.3(0.06) &  97.4(0.05) \\
                         &                       & $.2$  &   0.1(0.01) &  99.0(0.03) &  96.9(0.06) &  98.3(0.04) \\
                         &                       & $.5$  &   0.8(0.02) &  99.8(0.02) &  96.7(0.06) &  98.8(0.03) \\
                         \hline
\multirow{12}{*}{$1000$}  & \multirow{3}{*}{$.1$} & $.1$ &   2.3(0.03) &  99.4(0.02) &  97.8(0.05) &  99.8(0.01) \\ 
                         &                       & $.2$  &  -1.5(0.07) &  99.9(0.01) &  97.5(0.05) & 100.0(0.00) \\
                         &                       & $.5$  & -83.2(0.11) & 100.0(0.00) & 100.0(0.00) & 100.0(0.00) \\
                         & \multirow{3}{*}{$.5$} & $.1$  &   0.1(0.00) &  99.4(0.02) &  97.6(0.05) &  98.8(0.03) \\ 
                         &                       & $.2$  &   0.7(0.01) &  99.8(0.02) &  97.8(0.05) &  99.2(0.03) \\ 
                         &                       & $.5$  &   2.9(0.04) &  99.3(0.03) &  97.5(0.05) &  99.8(0.01) \\ 
                         & \multirow{3}{*}{$1$}  & $.1$  &   0.0(0.00) &  99.2(0.03) &  97.6(0.05) &  98.6(0.04) \\ 
                         &                       & $.2$  &   0.2(0.00) &  99.5(0.02) &  97.4(0.05) &  98.8(0.03) \\ 
                         &                       & $.5$  &   1.7(0.02) &  99.8(0.01) &  97.1(0.05) &  99.2(0.03) \\ 
                         & \multirow{3}{*}{$2$}  & $.1$  &   0.0(0.00) &  98.6(0.04) &  97.1(0.05) &  98.0(0.04) \\ 
                         &                       & $.2$  &   0.3(0.01) &  99.0(0.03) &  97.2(0.05) &  98.7(0.04) \\ 
                         &                       & $.5$  &   1.2(0.02) &  99.1(0.03) &  96.6(0.06) &  98.9(0.03) \\ 
                         & \multirow{3}{*}{$5$}  & $.1$  &   0.0(0.00) &  98.2(0.04) &  96.8(0.06) &  97.7(0.05) \\ 
                         &                       & $.2$  &   0.1(0.00) &  99.1(0.03) &  97.5(0.05) &  98.6(0.04) \\ 
                         &                       & $.5$  &   0.5(0.01) &  99.6(0.02) &  97.0(0.05) &  98.7(0.04) \\ 
                         
\end{tabular}
\vspace{.75em}
\caption{Empirical bias (EBias) of $\covbest$ and empirical coverage proportion (ECP) of $\covlestnormal,\covlestscore$, and $\covlestbayes$ ($1,000$ bootstrap samples) across $100,000$ simulations. 
Values in parentheses are empirical standard errors.}
\label{tab:simresultscovr}
\end{table}
}

{\renewcommand{\baselinestretch}{1}
\begin{table}[p]
\centering
\begin{tabular}{ccc|rr|rr}
$n$ & $s$ & $\sigma$ & Time $\covlestscore$ & Time $\covlestboot$ & Non-convergence ($\%$) & $\%$ $\covlestnormal<0$ \\
\hline
\multirow{12}{*}{$200$}  & \multirow{3}{*}{$.1$} & $.1$ & 1.01(0.00) & 32.1(0.05) &  8.4(0.09) & 99.5(0.02) \\
                         &                       & $.2$ & 1.02(0.00) & 32.8(0.06) & 26.4(0.16) & 99.9(0.01) \\
                         &                       & $.5$ & 1.02(0.00) & 33.9(0.07) & 41.8(0.20) & 00.0(0.01) \\
                         & \multirow{3}{*}{$.5$} & $.1$ & 1.01(0.00) & 31.7(0.05) &  0.1(0.01) & 15.6(0.11) \\
                         &                       & $.2$ & 1.01(0.00) & 31.9(0.05) &  1.0(0.03) & 62.6(0.15) \\
                         &                       & $.5$ & 1.02(0.00) & 32.2(0.05) & 13.4(0.12) & 99.7(0.02) \\
                         & \multirow{3}{*}{$1$}  & $.1$ & 1.01(0.00) & 31.8(0.05) &  0.0(0.00) &  3.7(0.06) \\
                         &                       & $.2$ & 1.01(0.00) & 31.9(0.05) &  0.2(0.02) & 21.1(0.13) \\
                         &                       & $.5$ & 1.01(0.00) & 32.1(0.05) &  3.9(0.06) & 98.6(0.04) \\
                         & \multirow{3}{*}{$2$}  & $.1$ & 1.01(0.00) & 32.1(0.05) &  0.0(0.00) &  0.7(0.03) \\
                         &                       & $.2$ & 1.01(0.00) & 32.1(0.05) &  0.1(0.01) & 10.5(0.10) \\
                         &                       & $.5$ & 1.01(0.00) & 32.1(0.05) &  1.9(0.04) & 78.8(0.13) \\
                         & \multirow{3}{*}{$5$}  & $.1$ & 1.01(0.00) & 32.3(0.05) &  0.1(0.01) &  0.4(0.02) \\
                         &                       & $.2$ & 1.01(0.00) & 32.4(0.05) &  0.7(0.03) &  8.1(0.09) \\
                         &                       & $.5$ & 1.01(0.00) & 32.5(0.05) &  4.2(0.06) & 67.8(0.15) \\
\hline
\multirow{12}{*}{$500$}  & \multirow{3}{*}{$.1$} & $.1$  & 1.01(0.00) & 31.9(0.05) &  2.2(0.05) & 82.3(0.12) \\
                         &                       & $.2$  & 1.02(0.00) & 32.3(0.05) & 13.8(0.12) & 97.8(0.05) \\
                         &                       & $.5$  & 1.02(0.00) & 33.9(0.06) & 34.5(0.19) & 99.7(0.02) \\
                         & \multirow{3}{*}{$.5$} & $.1$  & 1.01(0.00) & 31.7(0.05) &  0.0(0.00) &  3.4(0.06) \\
                         &                       & $.2$  & 1.01(0.00) & 31.7(0.05) &  0.2(0.01) & 19.5(0.13) \\
                         &                       & $.5$  & 1.01(0.00) & 31.9(0.05) &  4.1(0.06) & 93.3(0.08) \\
                         & \multirow{3}{*}{$1$}  & $.1$  & 1.01(0.00) & 31.9(0.05) &  0.0(0.00) &  0.6(0.02) \\
                         &                       & $.2$  & 1.01(0.00) & 31.7(0.05) &  0.0(0.01) &  6.1(0.08) \\
                         &                       & $.5$  & 1.01(0.00) & 31.9(0.05) &  1.0(0.03) & 48.0(0.16) \\
                         & \multirow{3}{*}{$2$}  & $.1$  & 1.01(0.00) & 31.9(0.05) &  0.0(0.00) &  0.2(0.01) \\
                         &                       & $.2$  & 1.01(0.00) & 31.9(0.05) &  0.0(0.01) &  1.9(0.04) \\
                         &                       & $.5$  & 1.01(0.00) & 31.9(0.05) &  0.6(0.02) & 22.8(0.13) \\
                         & \multirow{3}{*}{$5$}  & $.1$  & 1.01(0.00) & 32.2(0.05) &  0.1(0.01) &  0.1(0.01) \\
                         &                       & $.2$  & 1.01(0.00) & 32.3(0.05) &  0.3(0.02) &  1.0(0.03) \\
                         &                       & $.5$  & 1.01(0.00) & 32.3(0.05) &  1.8(0.04) & 21.7(0.13) \\
\hline
\multirow{12}{*}{$1000$}  & \multirow{3}{*}{$.1$} & $.1$ & 1.01(0.00) & 31.8(0.05) &  0.7(0.03) & 31.5(0.15) \\
                         &                       & $.2$  & 1.01(0.00) & 32.1(0.05) &  6.1(0.08) & 86.0(0.11) \\
                         &                       & $.5$  & 1.02(0.00) & 34.3(0.07) & 28.6(0.17) & 98.4(0.05) \\
                         & \multirow{3}{*}{$.5$} & $.1$  & 1.01(0.00) & 31.9(0.05) &  0.0(0.00) &  0.9(0.03) \\
                         &                       & $.2$  & 1.01(0.00) & 31.8(0.05) &  0.1(0.01) &  6.2(0.08) \\
                         &                       & $.5$  & 1.01(0.00) & 31.8(0.05) &  1.5(0.04) & 49.1(0.16) \\
                         & \multirow{3}{*}{$1$}  & $.1$  & 1.01(0.00) & 31.7(0.05) &  0.0(0.00) &  0.1(0.01) \\
                         &                       & $.2$  & 1.01(0.00) & 31.7(0.05) &  0.0(0.00) &  1.8(0.04) \\
                         &                       & $.5$  & 1.01(0.00) & 31.9(0.05) &  0.4(0.02) & 18.1(0.12) \\
                         & \multirow{3}{*}{$2$}  & $.1$  & 1.01(0.00) & 31.8(0.05) &  0.0(0.00) &  0.0(0.00) \\
                         &                       & $.2$  & 1.01(0.00) & 32.0(0.05) &  0.0(0.00) &  0.4(0.02) \\
                         &                       & $.5$  & 1.01(0.00) & 32.0(0.05) &  0.3(0.02) & 10.2(0.10) \\
                         & \multirow{3}{*}{$5$}  & $.1$  & 1.01(0.00) & 32.4(0.05) &  0.0(0.00) &  0.0(0.00) \\
                         &                       & $.2$  & 1.01(0.00) & 32.3(0.05) &  0.1(0.01) &  0.3(0.02) \\
                         &                       & $.5$  & 1.01(0.00) & 32.3(0.05) &  1.0(0.03) & 11.4(0.10) \\
\end{tabular}
\vspace{.75em}
\caption{Average and empirical standard error of computation time for $\covlestscore,\covlestboot$ relative to $\covlestnormal$, proportion of non-convergent $\covbest$, and proportion of $\covlestnormal<\covref$, for $100,000$ simulations.
}
\label{tab:simresultscomp}
\end{table}
}

Table \ref{tab:simresultscomp} summarizes computational aspects of the procedure. The score method is essentially just as fast as the Delta method, but with the advantage of not being able to return a computed lower bound that is less than $\covref$. In contrast, Table \ref{tab:simresultscomp} shows the percentage of simulations in each configuration that yielded an estimated $\covlestnormal<\covref$, and would therefore be unhelpful in practice. This happens in a substantial proportion of cases, especially at lower sample sizes and for flatter dose-response curves. This can be regarded as a disadvantage of the Delta method in this context, and a compelling reason to adopt $\covlestscore$ in its place. Further, our bootstrap lower limit $\covlestbayes$ based on $1,000$ samples takes only an order of magnitude more time than $\covlestnormal$ and $\covlestscore$. Overall, we recommend $\covlestscore$ based on a balance of computation time and empirical coverage, although we emphasize that it is plausible that both $\covlestscore$ and $\covlestbayes$ could be computed in typical applications.


%% file: 06-dataanalysis.tex

\section{Benchmark Dose Analysis of Prenatal Alcohol Exposure}\label{sec:paeanalysis}
We consider data from six longitudinal cohort studies
of alcohol consumption by pregnant women conducted in Detroit \citep{doi:10.1111/j.1530-0277.1993.tb00744.x}, Pittsburgh \citep{DAY1991329, Galearticle}, Atlanta \citep{article, Brownarticle}, and Seattle \citep{Streissguthetal1981}.
The research question of interest is to quantify levels of prenatal 
alcohol exposure (PAE) associated with the development of clinically 
important cognitive deficits in children. 
In these six cohort studies, children were followed from infancy through young adulthood and investigators administered a range of
neuropsychological tests to assess IQ and four domains of cognitive 
function: learning and memory, executive function, and academic achievement in reading and mathematics. 
To obtain an overall cognitive function score for each children, a structural equation model was fitted for each cohort. We then used the estimated cognitive function score as the outcome measure in the analyses that follow. 
For details on this approach, see Jacobson et al. (submitted) and \citet{akkaya_hocagil_benchmark_2023}.  

Data on maternal alcohol consumption were summarized in terms of average alcohol intake per day (ounces of absolute alcohol, (AA)/day) during pregnancy, and data on a broad range of potential confounders were collected in these cohorts. Since each cohort provided a somewhat different set of confounding variables, for these data \citet{jacobson_effects_2023} modeled the exposure variable, average alcohol intake per day, as a function of the potential confounders via linear regression model and use this to estimate a  propensity score \citep{rosenbaum83}. We used these propensity scores as covariates \citep{rosenbaum83,Imbens} in the following dose-response model.
 
Let $Y_i$ represent the cognitive function score, $\cov_i = \log(a_i+1)$ where $a_i$ is the average alcohol intake per day during pregnancy, $j_i\in\left\{1,\ldots,6\right\}$ index the cohort to which subject $i$ belongs, and $\concov_{ij} = s_{j_i}$ denote the computed propensity scores for subject $i=1,\ldots,2226$ in cohort $j_i$. 
A log-transformation was applied to $a_i$ to reduce the influence of very high exposure values on the fitted model.
The dose-response model is then

\begin{equation}\label{eqn:modelbspline}
    Y_{i} = \intercept +
   f(\cov_{i}) + 
    g_{j_i}(\concov_{ij}) +
    \responsesd\epsilon_i, \
    \epsilon_i\iid\responsedist, \ 
\end{equation}
where $f$ is an unknown, smooth monotone dose-response function, and $g_j,j=1,\ldots,6$ are unknown smooth functions. 
We apply the basis expansions of Section \ref{sec:monotonesplines} and fit the model using the methods described in Section \ref{sec:modelfitting}.   We compute the estimated BMD, $\covbest$, using the method of Section \ref{sec:numericalmethods}, and the three candidate BMDLs, $\covlestnormal$, $\covlestscore$, and $\covlestboot$, using the methods of Section \ref{sec:computationbmdl}.



\begin{figure}
\centering
\subfloat[Estimated dose-response curve, $\smoothfun$.]{
	\centering
	\includegraphics[width=4in]{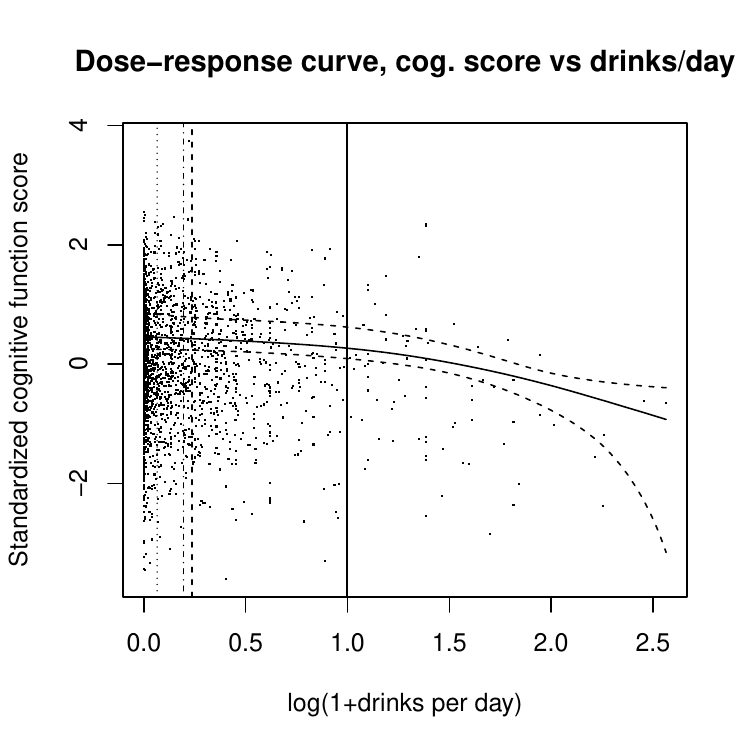}
}
\\
\subfloat[Posterior samples of the benchmark dose, $\covb$]{
	\centering
	\includegraphics[width=4in]{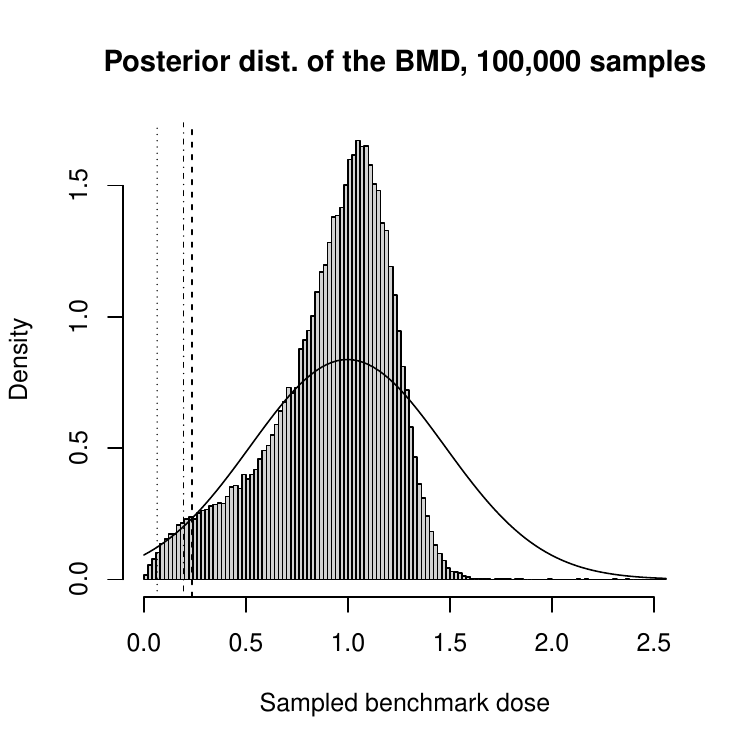}
}
\caption{(a) Estimated dose-response curve (---) and uncertainty bands
computed as pointwise $95\%$ credible intervals from $100,000$ samples from the posterior of the fitted curve (- - -). Vertical lines are the estimated BMD, $\covbest$ (---) and the BMDLs:
$\covlestnormal$ ($\cdots$), $\covlestscore$ (- - -), and $\covlestboot$ ($-\cdot-$). The dose-response curve is very flat,
and this yields high uncertainty in $\covbest$ which is
reflected in the relatively low values for the BMDs.
(b) $100,000$ samples from the approximate posterior distribution of $\covbest$ 
(\textcolor{gray}{$\blacksquare$}) and Normal approximation to
this distribution (---) upon which $\covlestnormal$ is based.
The approximation is not accurate, and this is reflected by $\covlestnormal$ being far from the $2.5\%$ percentile of this distribution, $\covlestboot$.
Observe that $\covlestscore$ is much closer to $\covlestboot$,
and is computed three orders of magnitude faster.
}
\label{fig:paedoseresponsecurve}
\end{figure}

Figure \ref{fig:paedoseresponsecurve} represents the  estimated dose-response curve obtained from the fitted monotone additive model along with the estimated BMD and corresponding BMDLs. 
The estimated BMD for $p_0=0.025$ and $\BMR=0.01$ 
was $\covbest=1.00$ and the three BMDLs were $\covlestscore = 0.236$, $\covlestboot = 0.195$, and $\covlestnormal = 0.065$. 
The Delta method estimate, $\covlestnormal$, is much closer to zero
than the other two, and hence communicates a much more
conservative clinical recommendation;
combined with its empirical coverage being too high in simulations,
we remark that this can be regarded as a disadvantage of the Delta
method over the two proposed BMDLs.
The conclusion based on $\covlestscore$ is that expectant mothers should consume not more than $\exp(0.236)-1 = 0.266$ oz of absolute alcohol which corresponds to approximately $0.45$ standard drinks per day, on average, to avoid clinically significant cognitive defects in their children.

The total amount of time that the procedure took to fit the model, estimate the BMD, and compute the three BMDLs based on $100,000$ bootstrap iterations was 396 seconds on a 2021 M1 Macbook Pro with $64$Gb of RAM. 
Model fitting and estimating the BMD took $3.93$ seconds 
and $30$ microseconds, respectively. 
The computation time for $100,000$ iterations of the parametric bootstrap for $\covbest$ was the longest at 390 seconds,
and produced a result close to $\covlestscore$ which was computed
nearly $1000\times$ faster in only $0.4$ seconds.
Without running the bootstrap, the procedure would have taken a total of $5.53$ seconds to perform a full semi-parametric benchmark dose analysis on these data.

%% file: 07-discussion.tex

\section{Discussion}\label{sec:discussion}

We recommend avoiding the Delta method ($\covlestnormal$; Section \ref{subsubsec:deltabmdl}) for forming confidence intervals for $\covbest$ in general.
The estimate $\covbest$ is a highly 
non-linear function of the MLE, and should not be expected to have a sampling distribution that
is close to Gaussian. 
An example of this is shown in Figure \ref{fig:paedoseresponsecurve} (b) for
the PAE data analysis, in which the approximate posterior distribution of $\covbest$ is extremely non-normal,
leading to a substantial overestimation of uncertainty and a low value of $\covlestnormal$.

The estimated dose-response curve shown in Figure \ref{fig:paedoseresponsecurve} (a) for the PAE data analysis is very flat at low exposure values.
This appears to be due to a large amount of noise relative to signal in the PAE data themselves,
suggesting that a single exposure may not be sufficiently informative \citep{jacobson_effects_2023}.
\citet{akkaya_hocagil_benchmark_2023} performs a benchmark dose analysis for PAE using a bivariate exposure that incorporates both frequency and severity of drinking.
In future work we plan to develop an efficient computational method for the semi-parametric bivariate exposure case,
making use of the fast computational methods developed in the
present manuscript to improve the practical application of benchmark dose estimation with bivariate exposure.

Although the present paper presents methodology that is specifically tied to computations related to benchmark dose estimation and its associated lower confidence limit, there is potential for the methods to apply in more general settings. In particular, it is useful to recognize that estimating a benchmark dose represents a specific example of a broader class of inverse estimation problems,
where the object of interest is a point on the $x$-axis in a semi-parametric regression model, and is hence obtained by solving a random nonlinear equation. 
Inverse problems arise in a wide range of applied settings, for example, chemical calibration or medical imaging, and it would be interesting to apply similar
approaches to those used in the present paper in other such problems.

%% file: appendix-02-algorithms.tex
\section{Algorithms}\label{app:algorithms}
{\renewcommand{\baselinestretch}{1}
\begin{algorithm}
\caption{de Boor's Algorithm for Spline Curve Evaluation}
\label{alg:deboor}
\begin{algorithmic}
\Require $\cov\in\R,\unconstrainedbasisweightvec=(\unconstrainedbasisweight_1,\ldots,\unconstrainedbasisweight_\numspline), p\in\N,\knots=(\knot_1,\ldots,\knot_{\numspline+p})$ such that $\smoothfun(\cov) = \sum_{l=1}^{\numspline}\basisvec_{l,p}(\cov)\unconstrainedbasisweight_l$.
\State Define $k : x\in[\knot_k,\knot_{k+1})$
\For{$j=0,\ldots,p-1$} $\unconstrainedbasisweight^{*}_j = \unconstrainedbasisweight_{j+k-(p-1)}$
\EndFor
\For{$r=1,\ldots,p-1$}
    \For{$j=(p-1),\ldots,r$}
        \State $\texttt{a} = (x - \knot_{j+k-(p-1)}) / (\knot_{j+1+k-r} - \knot_{j+k-(p-1)})$
        \State $\unconstrainedbasisweight^{*}_j = (1-\texttt{a})\unconstrainedbasisweight^{*}_{j-1} + \texttt{a}\unconstrainedbasisweight^{*}_j$
    \EndFor
\EndFor

\Return $\unconstrainedbasisweight^{*}_{p-1} = \smoothfun(\cov)$
\end{algorithmic}
\end{algorithm}
}
{\renewcommand{\baselinestretch}{1}
\begin{algorithm}
\caption{Reflective Newton Line-Search for Benchmark Dose Estimation}
\label{alg:newtonreflect}
\begin{algorithmic}
\Require $\constrainedbasisweightvecmle$ re-parameterized estimated spline weights; $\knots$ spline knots;
$\responsesdest>0,\bmrconst,\epsilon>0,\order=4,$
$\covref < \covmax$
\State Let $t=0,\cov^0 = (\covref+\covmax)/2,\texttt{u} = 1+\epsilon,\texttt{u}^\prime=1$.
\State $\texttt{f}_0 = \texttt{deBoor}(\covref,\constrainedbasisweightvecmle,\knots,\order)$
\State Let $\constrainedbasisweightvecmle^\prime = \left\{(\order-1)(\constrainedbasisweightmle_i - \constrainedbasisweightmle_{i-1})/(\knot_{i+\order-1}-\knot_{i})\right\}_{i=2}^{\numspline}$
\While{$|\texttt{u}| > \epsilon$}
    \State $\texttt{f} = \texttt{deBoor}(\cov^t,\constrainedbasisweightvecmle,\knots,\order)$,
    $\texttt{f}^\prime = \texttt{deBoor}(\cov^t,\constrainedbasisweightvecmle^\prime,\knots,\order-1)$
    \State $\texttt{u} = (\texttt{f}_0 - \texttt{f})/\responsesdest - \bmrconst$,
    $\texttt{u}^\prime = -\texttt{f}^\prime/\responsesdest$
    \State $\cov^{t+1} \gets \cov^t - \texttt{u} / \texttt{u}^\prime$
    \State $\cov^{t+1}\gets \texttt{Reflect}(\cov^{t+1},\covref,\covmax)$
    \State $t\gets t+1$
\EndWhile

\Return $\covbest = \cov^{t+1}$
\Require $\texttt{Reflect}(x,l,u) = \min(w,2(u-l)-w)+l$, where $w = |y-l|\text{mod}\{2(u-l)\}$.
\Require $\texttt{deBoor}(\cov,\boldsymbol{\beta},\knots,\order)$: Algorithm \ref{alg:deboor}
\end{algorithmic}
\end{algorithm}
}
{\renewcommand{\baselinestretch}{1}
\begin{algorithm}
\caption{Reflective Newton Line-Search for Benchmark Dose Lower Limit Computation}
\label{alg:newtonscore}
\begin{algorithmic}
\Require $\constrainedbasisweightvecmle$ re-parameterized estimated spline weights; $\knots$ spline knots;
$\responsesdest>0,\bmrconst,\epsilon>0,\order=4,$
$\covref < \covbest$;
$\betavar$ variance matrix; $0<\alpha<1, q: P(\chi^2_1<q)=\alpha$.
\State Let $t=0,\cov^0 = (\covref+\covbest)/2,\texttt{psi} = 1+\epsilon,\texttt{psi}^\prime=1$
\State Let $\constrainedbasisweightvecmle^\prime = \left\{(\order-1)(\constrainedbasisweightmle_i - \constrainedbasisweightmle_{i-1})/(\knot_{i+\order-1}-\knot_{i})\right\}_{i=2}^{\numspline}$
\While{$|\texttt{u}| > \epsilon$}
    \State $\texttt{f} = \texttt{deBoor}(\cov^t,\constrainedbasisweightvecmle,\knots,\order)$,
    $\texttt{f}^\prime = \texttt{deBoor}(\cov^t,\constrainedbasisweightvecmle^\prime,\knots,\order-1)$
    \State $\texttt{u} = (\texttt{f}_0 - \texttt{f})/\responsesdest - \bmrconst$,
    $\texttt{u}^\prime = -\texttt{f}^\prime/\responsesdest$,
    $\texttt{v} = \Unvar(x^t),\texttt{v}^\prime = \Unvar^\prime(x^t)$
    \State $\texttt{psi} = \texttt{u}^2 - \texttt{v}q,\texttt{psi}^\prime = 2\texttt{uu}^\prime - \texttt{v}^\prime q$
    \State $\cov^{t+1} \gets \cov^t - \texttt{psi} / \texttt{psi}^\prime$
    \State $\cov^{t+1}\gets \texttt{Reflect}(\cov^{t+1},\covref,\covbest)$
    \State $t\gets t+1$
\EndWhile

\Return $\covlestscore = \cov^{t+1}$
\Require $\texttt{Reflect}(x,l,u) = \min(w,2(u-l)-w)+l$, where $w = |y-l|\text{mod}\{2(u-l)\}$.
\Require $\texttt{deBoor}(\cov,\boldsymbol{\beta},\knots,\order)$: Algorithm \ref{alg:deboor}.
\end{algorithmic}
\end{algorithm}
}